\documentclass[twocolumn,showpacs,preprintnumbers,amsmath,amssymb]{revtex4}
\usepackage{graphicx}
\usepackage{dcolumn}
\usepackage{bm}

\begin{document}
 
\title{Plurality Voting: the statistical laws of democracy in Brazil}

\author{Luis E. Araripe, Raimundo N. Costa Filho, Hans J. Herrmann,
and Jos\'e S. Andrade Jr.}

\affiliation{Departamento de F\'{\i}sica, Universidade Federal
do Cear\'a, 60451-970 Fortaleza, Cear\'a, Brazil}

\date{\today}

\begin{abstract}
We explore the statistical laws behind the plurality voting system by
investigating the election results for mayor held in Brazil in
2004. Our analysis indicate that the vote partition among mayor
candidates of the same city tends to be ``polarized'' between two
candidates, a phenomenon that can be closely described by means of a
simple fragmentation model. Complex concepts like ``government
continuity'' and ``useful vote'' can be identified and even
statistically quantified through our approach.
\end{abstract}


\maketitle 

Understanding the process by which the individuals of a society make
up their minds and reach opinions about different issues can be of
great importance. In this context, the election is a fundamental
democratic process and the vote certainly represents the most
effective instrument for regular citizens to promote significant
changes in their communities. 

General elections in Brazil are held every four years, when citizens
vote for executive as well as legislative mandates. Voting is
compulsory and ballots are collected in electronic voting machines.
In the case of the legislative mandates, which include elections for
congressmen, state deputies and city counselors, a {\it proportional}
voting system is used, where candidates compete for a limited number
of seats and become elected in proportion to their corresponding
voting fraction. In two previous studies \cite{Costa98,Costa03}, the
statistical analysis for the Brazilian 1998 and 2002 elections
revealed that the proportional voting process for federal and state
deputies displays scale invariance \cite{Ball04}. It has been shown
that the distribution of the number of candidates $N$ receiving a
fraction of votes $v$ follows a power-law $N(v)\sim v^\alpha$, where
$\alpha\approx-1$, extending over two orders of magnitude. The
striking similarity in the distribution of votes in all states,
regardless of large diversities in social and economical conditions in
different regions of the country, has been taken as an indication of a
common mechanism in the decision process. More precisely, it has been
suggested that the explanation for this robust scale invariance is a
multiplicative process in which the voter's choice for a candidate is
governed by a product, instead of a sum of probabilities
\cite{West89}.

For the selection to executive mandates (president, state governors
and mayors), one of the most common election formats is the so called
{\it plurality} voting system, where the winning candidate is only required
to receive the largest number of votes in his/her favor, after which
all other runners automatically and completely lose. This system is
applied in 43 of the 191 countries in the United Nations, Brazil being
the largest democracy in this group. Plurality voting has been
studied extensively in political science~\cite{Norris} and effects
such as the approval of previous administrations or tactical voting,
including the so called ``useful vote'', have been discussed from
psychological and sociological points of view. What is missing is a
careful statistical description and a mathematical model that include
these effects revealing common mechanisms in the decision process.

Here we analyse for the first time the election statistics for an
executive mandate in Brazil. On 6 October 2004, there was an election
in Brazil's 5,562 cities in which 102,817,864 electors chose one from
among up to 14 candidates for mayor. The collection of ballots was
entirely electronic, thus permitting a very rapid count and
publication of the results \cite{TSE}. In Fig.~1 we show the
distribution of the fraction of votes $v$ for the winner (right) and
for the loser (left), if only two candidates were in the race.  The
superposition of both sides leads to a distribution that displays a
pronounced cusp at $v=0.5$ and differs strongly from a uniform
distribution. As shown in Fig.~1 (solid lines), the entire data set
can be well described by an exponential decay of the form,
\begin{equation}\label{Eq_dist}
P(v) \propto exp\left(\frac{-|v-0.5|}{\lambda}\right)~,
\end{equation}
with the parameter $\lambda \approx 0.08$. The values for the left and
right sides correspond to the excess and deficit of votes for the
winning and losing candidates, respectively. The sharpness of the
curve beautifully illustrates the effect of polarization which drives
a typical ballot close to the marginal situation of a tie.

In Fig.~2 we show the statistics of the winner for cities with three
and four candidates. As depicted, both distributions display a
cusp-shaped maximum close to the same value $v=0.5$, and exponential
tails on both sides. This behavior can be described by a
generalization of the fragmentation model of Ref.~\cite{Derrida87},
based on the well known fact that the approval (or disapproval) of the
previous municipal administration usually decides rather early whether
the acting mayor or the candidate he/she supports as follower is
reelected or not. We start by dividing the electorate into two
fractions, $v_{1}$ and $r_{1}=1-v_{1}$. Keeping intact the fraction
$v_{1}$, we divide $r_{1}$ into $v_{2}$ and $r_{2}=r_{1}-v_{2}$. At a
third step, while $v_{2}$ remains undivided, the fraction $r_{2}$ is
partitioned again, and so on. As opposed to Ref.~\cite{Derrida87},
where the limit of an infinite number of fragments is investigated,
here we consider a process in which a finite number $n$ of fragments
(fraction of the electorate) is generated with sizes that can be
written as $v_{1}=x_{1}$, $v_{2}=(1-x_{1})x_{2}$,
$v_{3}=(1-x_{1})(1-x_{2})x_{3}$, $...$,
$v_{n-1}=(1-x_{1})(1-x_{2})...(1-x_{n-1})x_{n}$, and
$v_{n}=(1-x_{1})(1-x_{2})...(1-x_{n-2})(1-x_{n-1})$, with $0 \le x \le
1$ being a random variable distributed according to the same function
$\rho(x)$. This randomness excludes at this point any tactical voting
strategies. We simply attribute to each candidate $i$ a fraction of
votes $v_{i}$, with $i=1$, $2$, $3$, ..., $n-1$, $n$. This is
justifiable if we assume that $v_{i}$ should be closely related with
the fraction of electors ``decided'' to vote in candidate $i$. In this
way, it is reasonable to adopt the distribution given by
Eq.~(\ref{Eq_dist}) (see also Fig.~1) as a first approximation for
$\rho(x)$. Under this framework, one can also think of $x_{1}$ as
being the fraction of the electorate voting for ``continuity''. 
Following this model, for example, the fraction of votes $v_{max}$ 
of the most voted among a finite number $n$ of competing candidates 
is given by,
\begin{eqnarray}\label{Eq_vmax}
v_{max}=\max[v_{1}, v_{2}, v_{3}, ..., v_{n-1}, v_{n}]~.
\end{eqnarray}
For the numerical solution of our model, we first generate $n-1$
random numbers distributed according to Eq.~(\ref{Eq_dist}). From
these, we calculate the entire set of $v_{i}$ fractions and determine
the largest one, $v_{max}$, the second largest one, $v_{2max}$, the
third largest one, $v_{3max}$, and so on. We repeat this process
$N=10^{5}$ times in order to produce histograms for $v_{max}$,
$v_{2max}$ and $v_{3max}$, as displayed in Figs.~2, 4a and 4b,
respectively. As shown in Fig.~2, the agreement between the real data
and the model predictions for $v_{max}$ with $n=2$ and $3$ is very
good, without the need of any adjusting parameter of the fragmentation
model. This confirms the validity of our approach. 

The selection of one among $n$ candidates by the population during an
electoral campaign is certainly not a static process. For example, the
dynamics of a typical voting process that is studied here is shown in
Fig.~3, where we present the results of a sequence of polls made
before the mayor election in S\~ao Paulo during the campaign of 2004
\cite{IBOPE}. First, the time evolution of these polls illustrates
well the ``polarization'' between the first and the second most voted
candidates. Second, the growth in popularity of both candidacies is a
clear consequence of the loss of votes of the two less voted
candidates. This tactical transference of votes is explained as
follows. Being driven by the results of election polls widely spread
in the media during the campaign, the electors tend to adopt the
so-called ``useful vote'', either to try to guarantee or to prevent
the victory in the first round of the most voted candidate.

As shown in Fig.~4a, the fraction of votes for the second most voted
among three candidates ($n=3$) clearly reveals a tendency towards a
peak that is also surprisingly close to $v=0.5$. Interestingly, this
result only agrees with the model prediction if we admit a shift to
the right, i.e. a systematic excess of votes for the second candidate
which is due to the ``useful vote'' effect. The results shown in
Fig.~4b confirm the hypothesis of a polarized election. Due to the
``useful vote'', a significant number of votes is transferred from the
third to the other candidates.

In summary, based on a Brazilian dataset for plurality voting of
unseen quality, we have discovered the existence of a strongly peaked
exponential distribution in the case of only two candidates, showing
the effect of polarization. When more candidates are involved, a
simple fragmentation model based on early decisions concerning
continuity is able to explain the shape of the distribution of the
winner. Taking into account the polarization due to the ``useful
vote'', we can justify a systematic shift in the distribution of the
losing candidates.

We thank Josu\'{e} Mendes Filho and Andr\'e Moreira for discussions
and CNPq, CAPES, FUNCAP and the Max-Planck prize for financial
support.

\begin{figure}
\includegraphics[width=8cm]{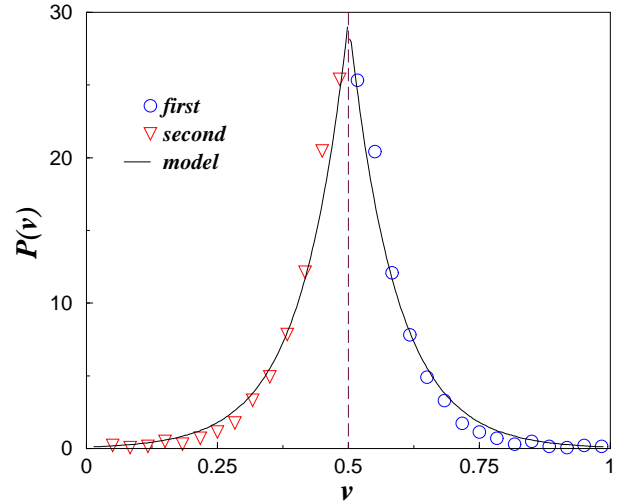}
\caption{Histograms of voting fraction for mayor elections of
Brazilian cities in 2004. The data correspond to elections with only
two candidates and all ordinates have been divided by a factor of
$10^{3}$. The circles give the frequency of fraction votes for the
winner, while the downward triangles are the results for the
loser. The solid lines are symmetric with respect to $v=0.5$ and
represent the best fits to the data by the exponential function,
$P(v) \propto \exp(-|v-0.5|/\lambda)$.}
\end{figure}

\begin{figure}
\includegraphics[width=8cm]{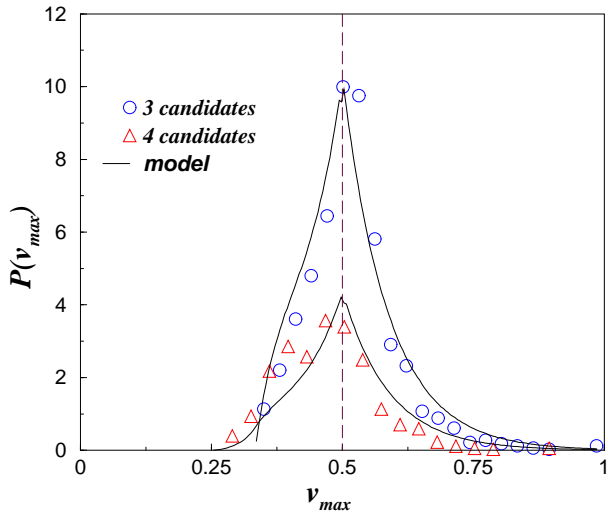}
\caption{Histograms of the voting fraction for the most voted candidates
$v_{max}$ in elections with three (circles) and four (upward
triangles) candidates. The solid lines are the predictions of the
fragmentation model.}
\end{figure}

\begin{figure}
\includegraphics[width=8cm]{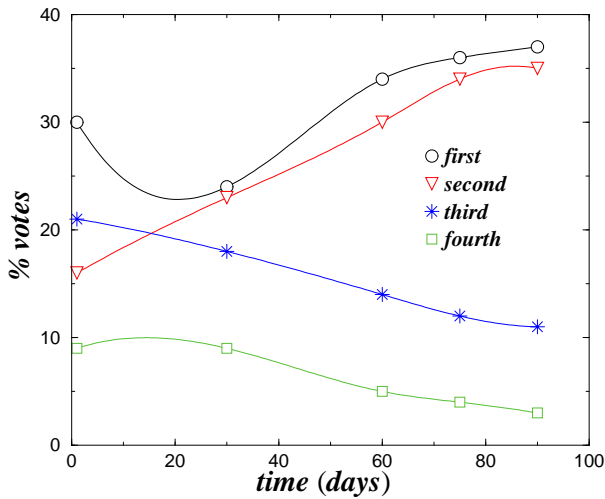}
\caption{Time evolution of the percentage of votes for the first 
four most voted candidates participating in the election for mayor in
S\~ao Paulo during the campaign of 2004. These polls have been made by
the Brazilian agency IBOPE \cite{IBOPE}. The time in days is counted
from the date of the first poll, namely 28 of June of 2004. The gradual
approximation between the first two most voted candidates shows the
polarization phenomenon, while the growth of both candidacies
illustrates the ``useful vote'' effect. The solid lines are cubic
splines drawn to facilitate the view.}
\end{figure}

\begin{figure}
\includegraphics[width=8cm]{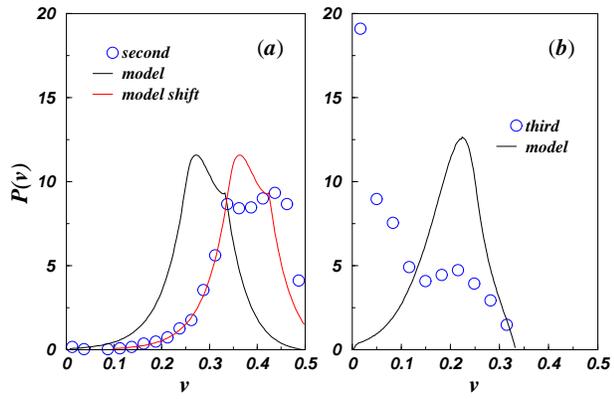}
\caption{Statistics of the second and third voted candidates in elections with
only three candidates. In (a) we see that the data for the second
voted candidate significantly deviate from the model predictions
(solid line), revealing the effect of the ``useful vote''. The full
line is obtained by shifting the model results by about 0.1. The
results for the third voted candidate in (b) show an opposite
deviation between data (circles) and model predictions (solid line).}
\end{figure}

\end{document}